\documentclass[12pt]{article}
\usepackage{amsmath}
\usepackage{amsfonts}
\usepackage{amssymb}
\usepackage{graphicx}

\input{tcilatex}
\begin{document}

\begin{titlepage}
\ \\
\begin{center}
\LARGE
{\bf
A Fundamental Lower Bound of 
Actuating Energy \\
for Broadband  Photon Switching
}
\end{center}
\ \\
\begin{center}
\large{
Masahiro Hotta
}\\
\ \\
\ \\
{\it
Department of Physics, Faculty of Science, Tohoku University,\\
Sendai, 980-8578, Japan\\
hotta@tuhep.phys.tohoku.ac.jp
}
\end{center}
\begin{abstract}
We derive a universal 
lower bound of actuating energy $E_{s}$ for  broadband  photon switching by using an uncertainty relation between time and the negative energy density of quantum fields. We find that broadband photon switching 
 between  perfect reflection and perfect transmission over a time $t_s$ should satisfy 
$E_{s}\geq\frac{\hbar}{6\pi t_s}$. 
\end{abstract}
\end{titlepage}

\bigskip

\section{Introduction}

\ \newline

Considerable effort has been made to realize a network of long-distance
quantum communication by photons. If the range of available wavelengths is
very broad, more information can be sent on a quantum communication network
using encoding strategies than by monochromatic-beam schemes. For network
applications, the technology of photon switching \cite{HY} (or similarly,
switchable mirrors \cite{H}) are expected to play an important role. From
the technological point of view, amount of energy required to operate photon
switches should be as small as possible. The energy to actuate the switch
depends on the implementation method and, in principle, various kinds of
implementation of photon switching may be considered. Then a nontrivial
question arises. How can we reduce the actuating energy $E_{s}$ of photon
switching in a fundamental level ? In this paper, we consider switching
devices, which take an actuating time $t_{s}$, designed to switch between
perfect reflection and perfect transmission for incident broadband photons
and give a fundamental lower bound of $E_{s}$. During the switching process,
the switch body emits undesired photons inevitably due to interactions
between the device and the electromagnetic field. The generation of photons
by changing boundary conditions is a familiar mechanism in quantum field
theory, which cause, for example, the dynamical Casimir effect\cite{dc}. We
evaluate the minimum work done by the switch, utilizing a gedanken
experiment based on a uncertainty relation between time and the negative
energy of multi-particle states. Of course, the actuating energy $E_{s}$ is
bounded below by the work to generate the undesired photons. Using the
evaluation of the work, we show that $E_{s}\geq \frac{\hbar }{6\pi t_{s}}$.
It should be emphasized that the lower bound is satisfied for arbitrary
broadband photon switching between perfect penetration and perfect
reflection with actuating time $t_{s}$. In this paper, we assume that the
photon beams propagate in vacuum and the bandwidth is so broad that
arbitrary spatial configuration of electromagnetic field can be realized
with a good precision.

\section{Analysis of modes propagating along the x axis}

\ \newline

\bigskip As well known, free electromagnetic field can be expanded as%
\begin{equation*}
A^{\mu }=\int \int \int \sqrt{\frac{\hbar }{\left( 2\pi \right)
^{3}2\left\vert \mathbf{k}\right\vert }}\sum_{\lambda }e_{\lambda }^{\mu
}\left( \mathbf{k}\right) \left[ a^{\lambda }\left( \mathbf{k}\right)
e^{i\left( \mathbf{kx-}\left\vert \mathbf{k}\right\vert t\right)
}+a^{\lambda \dag }\left( \mathbf{k}\right) e^{-i\left( \mathbf{kx-}%
\left\vert \mathbf{k}\right\vert t\right) }\right] d^{3}k,
\end{equation*}%
where we set the light velocity equal to one, $e_{\lambda }^{\mu }\left( 
\mathbf{k}\right) $ is polarization vectors and $a^{\lambda }\left( \mathbf{k%
}\right) ,~a^{\lambda \dag }\left( \mathbf{k}\right) $ are annihilation and
creation operators. In principle, all modes can be coupled to and excited by
photon switches during switching operations. The actuating energy $E_{s}$ of
a photon switch is bounded below by a sum of inevitable works for all modes
to generate undesired excitations. Among the modes, let us concentrate on
analysis of the modes which propagate along the $x$ axis because we put the
mirror boundary of the switch at $x=0$ in the $yz$ plane. \ Of course, $%
E_{s} $ is also bounded below by the work to produce undesired photons of
the $x$-axis modes. A beam propagating along the $x$ axis with its
crosssection $B$ is described \cite{g} by

\begin{equation*}
A^{\mu }(x,t)=\frac{1}{\sqrt{B}}\sum_{\lambda }e_{\lambda }^{\mu }\phi
^{\lambda }(x,t).
\end{equation*}%
Here $B$ is set large enough. The constant vector $e_{\lambda }^{\mu }$
denotes polarizations of photon and $\lambda $ takes two values of the
polarization. The massless fields $\phi ^{\lambda }$ satisfy the equation of
motion given by

\begin{equation}
\left[ \frac{\partial ^{2}}{\partial t^{2}}-\frac{\partial ^{2}}{\partial
x^{2}}\right] \phi ^{\lambda }(t,x)=0.  \label{eqm}
\end{equation}%
For later convenience, we suppress the superscript $\lambda $ of $\phi
^{\lambda }$. It should be noted that $\phi $ has a residual electromagnetic
gauge symmetry defined by $\phi \rightarrow \phi +const.$. The general
solution of Eq.\ (\ref{eqm}) is written as a sum of left- and right-moving
components: $\phi (x,t)=\phi _{+}\left( x^{+}\right) +\phi _{-}\left(
x^{-}\right) $, where $\phi _{+}\left( x^{+}\right) $ denotes the
left-moving field and $\phi _{-}\left( x^{-}\right) $ the right-moving field
with light-cone coordinates $x^{\pm }=t\pm x$. Note that the usual
plane-wave modes give a complete and orthogonal basis. \ In the left-moving
field system, for example, an arbitrary function $\ F(x^{+})~$of $x^{+}$ can
be uniquely expanded by use of the modes defined by

\begin{equation}
u_{\omega }(x_{+})=\sqrt{\frac{\hbar }{4\pi \omega }}e^{-i\omega
x^{+}},~(\omega \geq 0)  \label{p-mode}
\end{equation}%
as follows.

\begin{equation*}
F(x^{+})=\int_{0}^{\infty }d\omega \left[ F_{\omega }u_{\omega
}(x^{+})+F_{\omega }^{\ast }u_{\omega }^{\ast }(x^{+})\right] .
\end{equation*}%
The mode functions are orthogonal to each other in terms of the norm defined
by

\begin{equation*}
\left( f,g\right) =\frac{i}{\hbar }\int_{-\infty }^{\infty }\left[ f^{\ast
}(x^{+})\partial _{+}g(x^{+})-\partial _{+}f^{\ast }(x^{+})g(x^{+})\right]
dx^{+}.
\end{equation*}%
You can check directly the orthonormality such that

\begin{eqnarray}
\left( u_{\omega },~u_{\omega ^{\prime }}\right) &=&\delta \left( \omega
-\omega ^{\prime }\right) ,  \label{n1} \\
\left( u_{\omega }^{\ast },~u_{\omega ^{\prime }}\right) &=&0, \\
\left( u_{\omega }^{\ast },~u_{\omega ^{\prime }}^{\ast }\right) &=&\delta
\left( \omega -\omega ^{\prime }\right) .  \label{n3}
\end{eqnarray}
Due to this fact, each field is expressed by the plane-wave expansion:%
\begin{equation*}
\phi _{d}=\int_{0}^{\infty }d\omega \sqrt{\frac{\hbar }{4\pi \omega }}\left[
a_{\omega }^{d}e^{-i\omega x^{d}}+a_{\omega }^{d\dag }e^{i\omega x^{d}}%
\right] ,
\end{equation*}%
where $d=+$ or $-$. The creation and annihilation operators $a_{\omega
}^{d\dag }$ and $a_{\omega }^{d}$ obey the standard commutation relations
given by $\left[ a_{\omega }^{d},~a_{\omega ^{\prime }}^{d^{\prime }\dag }%
\right] =\delta ^{dd^{\prime }}\delta \left( \omega -\omega ^{\prime
}\right) $ because of Eq.(\ref{n1})-Eq.(\ref{n3}). By using the annihilation
operators, the normalized vacuum state $|0\rangle $ is defined by $a_{\omega
}^{d}|0\rangle =0$.

\bigskip

\section{Uncertainty relation between time and negative energy density of
quantum field}

\ \newline

The energy--momentum tensor $T_{\mu \nu }$ of the quantum field is defined
by adopting the normal operator ordering of $a_{\omega }^{d\dag }$ and $%
a_{\omega }^{d}$ in the classical expression of the tensor. It is remarkable
that the quantum interference effect between multi-particle states is able
to suppress quantum fluctuation of the field and to yield negative energy
density of the field \cite{BD}. For example, even though the classical
energy flux $\left[ \partial _{+}\phi _{+}(x^{+})\right] ^{2}$ of the
left-moving field is non-negative, the expectation value of the
corresponding quantum flux operator $T_{++}\left( x^{+}\right) =:\partial
_{+}\phi _{L}(x^{+})\partial _{+}\phi _{L}(x^{+}):$ can be negative. In
spite of the negative energy density, the expectation value of the total
energy flux $\int_{-\infty }^{\infty }T_{++}(x^{+})dx^{+}$ for an arbitrary
state remains non-negative because the total flux is given by $%
\int_{0}^{\infty }\hbar \omega a_{\omega }^{+\dag }a_{\omega }^{+}d\omega $.

By taking an arbitrary monotonically increasing $C^{1}~$function $f(x)$of $%
x\in \left( -\infty ,\infty \right) $satisfying $f(\pm \infty )=\pm \infty $%
, a new set of mode functions

\begin{equation}
v_{\omega }(x)=\sqrt{\frac{\hbar }{4\pi \omega }}e^{-i\omega f(x)},~(\omega
\geq 0)  \label{vm}
\end{equation}%
is obtained that can uniquely expand the field. Actually, we can show that
an aribitary function $g(x)$ can be uniquely expanded by the new mode
functions. Let us define a function given by%
\begin{equation*}
G(x)=g(f^{-1}(x)),
\end{equation*}%
where $f^{-1}$ is the inverse function of $f$. Then, by the Fourier
transformation, $G(x)$ is expressed as

\begin{equation*}
G(x)=\int_{-\infty }^{\infty }dk\tilde{G}(k)\sqrt{\frac{\hbar }{4\pi |k|}}%
e^{-ikx},
\end{equation*}%
where $\tilde{G}(k)$ is uniquely determined by $g(x)$ as follows.

\begin{eqnarray*}
\tilde{G}(k) &=&\sqrt{\frac{|k|}{\pi \hbar }}\int_{-\infty }^{\infty
}dxG(x)e^{ikx} \\
&=&\sqrt{\frac{|k|}{\pi \hbar }}\int_{-\infty }^{\infty
}dxg(f^{-1}(x))e^{ikx}.
\end{eqnarray*}%
Hence we obtain

\begin{eqnarray*}
G(x) &=&\int_{0}^{\infty }dk\tilde{G}(k)\sqrt{\frac{\hbar }{4\pi k}}%
e^{-ikx}+\int_{-\infty }^{0}dk\tilde{G}(k)\sqrt{\frac{\hbar }{4\pi (-k)}}%
e^{-ikx} \\
&=&\int_{0}^{\infty }d\omega \left[ \tilde{G}(\omega )\sqrt{\frac{\hbar }{%
4\pi \omega }}e^{-i\omega x}+\tilde{G}(-\omega )\sqrt{\frac{\hbar }{4\pi
\omega }}e^{i\omega x}\right] .
\end{eqnarray*}%
Substitution of  $f(x)$ into $G$ yields the following result which we really
want.

\begin{eqnarray*}
g(x) &=&G(f(x)) \\
&=&\int_{0}^{\infty }d\omega \left[ \tilde{G}(\omega )\sqrt{\frac{\hbar }{%
4\pi \omega }}e^{-i\omega f(x)}+\tilde{G}(-\omega )\sqrt{\frac{\hbar }{4\pi
\omega }}e^{i\omega f(x)}\right]  \\
&=&\int_{0}^{\infty }d\omega \left[ \tilde{G}(\omega )v_{\omega }(x)+\tilde{G%
}(-\omega )v_{\omega }^{\ast }(x)\right] .
\end{eqnarray*}%
The orthonormality in terms of the normal product can be also derived
straightforwardly. For example, we can calculate $\left( v_{\omega
},~v_{\omega ^{\prime }}\right) $  and obtain a right result as follows.  
\begin{eqnarray}
\left( v_{\omega },~v_{\omega ^{\prime }}\right)  &=&\frac{i}{\hbar }%
\int_{-\infty }^{\infty }\left[ v_{\omega }^{\ast }\partial v_{\omega
^{\prime }}-\partial v_{\omega }^{\ast }v_{\omega ^{\prime }}\right] dx 
\notag \\
&=&\frac{1}{2\pi }\int_{-\infty }^{\infty }e^{i\left( \omega -\omega
^{\prime }\right) f(x)}\frac{df}{dx}dx  \notag \\
&=&\frac{1}{2\pi }\int_{-\infty }^{\infty }e^{i\left( \omega -\omega
^{\prime }\right) x^{\prime }}dx^{\prime }  \notag \\
&=&\delta \left( \omega -\omega ^{\prime }\right) .  \label{n'1}
\end{eqnarray}%
In moving from the second line to the third line, we have used the
coordinate transformation $x^{\prime }=f(x)$. Similarly, it is proven that 
\begin{eqnarray}
\left( v_{\omega }^{\ast },~v_{\omega ^{\prime }}\right)  &=&0,  \label{n'2}
\\
\left( v_{\omega }^{\ast },~v_{\omega ^{\prime }}^{\ast }\right)  &=&\delta
\left( \omega -\omega ^{\prime }\right) .  \label{n'3}
\end{eqnarray}%
For the left-moving field $\phi _{+}$, the new expansion is given by

\begin{equation}
\phi _{+}\left( x^{+}\right) =\int_{0}^{\infty }d\omega \left[ b_{\omega
}^{+}v_{\omega }(x^{+})+b_{\omega }^{+\dag }v_{\omega }^{\ast }(x^{+})\right]
.  \label{exp}
\end{equation}%
Because of Eq.(\ref{n'1})-Eq.(\ref{n'3}), $b_{\omega }^{+\dag },~b_{\omega
}^{+}$ are new creation and annihilation operators which satisfy $\left[
b_{\omega }^{+},~b_{\omega ^{\prime }}^{+\dag }\right] =\delta \left( \omega
-\omega ^{\prime }\right) $ and depend linearly on the operators $a_{\omega
}^{+\dag }$ and $a_{\omega }^{+}$. A normalized quantum state $|\Phi \rangle 
$ defined by $b_{\omega }^{+}|\Phi \rangle =0$ is a squeezed state. For an
arbitrary $|\Phi \rangle $, the expectation value of the energy flux of $%
\phi _{+}$ is evaluated as

\begin{equation}
\langle \Phi |T_{++}\left( x^{+}\right) |\Phi \rangle =-\frac{\hbar }{24\pi }%
\left[ \frac{\dddot{f}\left( x^{+}\right) }{\dot{f}\left( x^{+}\right) }-%
\frac{3}{2}\left( \frac{\ddot{f}\left( x^{+}\right) }{\dot{f}\left(
x^{+}\right) }\right) ^{2}\right] ,  \label{f}
\end{equation}%
where the dot means a derivative in terms of $x^{+}$. The formula in Eq.(\ref%
{f}) can be given in several ways\cite{BD}. \ One simple way is the point
splitting method \cite{Ford} which is based on the relation:%
\begin{equation}
\langle \Phi |T_{++}\left( x^{+}\right) |\Phi \rangle =\lim_{\delta
\rightarrow 0}\left[ 
\begin{array}{c}
\langle \Phi |\partial _{+}\phi _{L}(x^{+}+\delta )\partial _{+}\phi
_{L}(x^{+})|\Phi \rangle \\ 
-\langle 0|\partial _{+}\phi _{L}(x^{+}+\delta )\partial _{+}\phi
_{L}(x^{+})|0\rangle%
\end{array}%
\right] .  \label{ps}
\end{equation}%
By substituting the field expansion in Eq.(\ref{exp}) into the first term of
the right-hand-side of Eq.(\ref{ps}), we obtain the result in Eq.(\ref{f}).

\bigskip

An important example of negative energy flux is generated by a monotonically
increasing $C^{1}$ function $f_{\varepsilon }\left( x\right) $ given by

\begin{align}
f_{\varepsilon }\left( x\right) & =\Theta \left( x_{i}-x\right) x  \notag \\
& +\Theta \left( x_{f}-x\right) \Theta \left( x-x_{i}\right) \left[ x_{i}-%
\frac{1}{\sqrt{\varepsilon }}+\frac{1}{\sqrt{\varepsilon }-\varepsilon
\left( x-x_{i}\right) }\right]  \notag \\
& +\Theta \left( x-x_{f}\right) \left[ \frac{\varepsilon }{\left( \sqrt{%
\varepsilon }-\varepsilon \left( x_{f}-x_{i}\right) \right) ^{2}}%
(x-x_{f})+x_{i}-\frac{1}{\sqrt{\varepsilon }}+\frac{1}{\sqrt{\varepsilon }%
-\varepsilon \left( x_{f}-x_{i}\right) }\right] ,  \label{f3}
\end{align}%
where $x_{i}\leq x_{f}$, $\Theta \left( x\right) $ is the step function and $%
\varepsilon =\left( \frac{12\pi \left\vert E_{n}\right\vert }{\hbar }\right)
^{2}$ is a nonnegative constant. For the squeezed state $|\Phi
_{shock}\rangle $ \ corresponding to $f_{\varepsilon }\left( x\right) $, \
the expectation value of the left-moving energy flux is computed as

\begin{equation}
\langle \Phi _{shock}|T_{++}(x^{+})|\Phi _{shock}\rangle =-\left\vert
E_{n}\right\vert \delta (x^{+}-x_{i})+\frac{\left\vert E_{n}\right\vert }{1-%
\frac{12\pi }{\hbar }\left\vert E_{n}\right\vert l}\delta \left(
x^{+}-x_{f}\right) ,  \label{n+p}
\end{equation}%
where $l=x_{f}-x_{i}(>0)$. The first term of the right-hand side shows the
flux of a shock wave with negative energy $-\left\vert E_{n}\right\vert $.
Because $\int_{-\infty }^{\infty }\langle \Phi _{shock}|T_{++}\left(
x\right) |\Phi _{shock}\rangle dx$ is positive, \ we obtain%
\begin{equation}
\frac{\left\vert E_{n}\right\vert ^{2}l}{\frac{\hbar }{12\pi }-\left\vert
E_{n}\right\vert l}\geq 0.
\end{equation}%
Because the numerator is definitely positive, the denominator must be
nonnegative, which leads to the inequality 
\begin{equation}
\left\vert E_{n}\right\vert \leq \frac{\hbar }{12\pi l}=\frac{\hbar }{12\pi
\left( x_{f}-x_{i}\right) }.  \label{nieq}
\end{equation}%
The length $l$ is the arrival interval between two shock waves with negative
energy and positive energy. Thus, Eq.\ (\ref{nieq}) shows an uncertainty
relation between time and the negative energy of the quantum field.

\bigskip

In the above argument, one might worry about whether nonanalytic behavior of
the step function which appears in $f_{\varepsilon }\left( x\right) $ leads
to a unphysical result or not. However there is no need to worry. In real
situations, we can always consider a smooth function $f_{\Lambda
,\varepsilon }\left( x\right) $ which depends on a physical cutoff parameter 
$\Lambda $ and approaches $f_{\varepsilon }\left( x\right) $ when $\Lambda $
becomes large. Because the bandwidth is assumed so broad that the cutoff $%
\Lambda $ is big enough, the expression of energy flux in Eq.(\ref{n+p}) can
be realized in a good precision.

\bigskip

\section{Local fluctuation equivalence between flux-vanishing \ states}

\ \newline

Next we mention a local fluctuation property of flux-vanishing states. The
general form of the function $f(x)$ which satisfies $\langle \Phi
|T_{++}|\Phi \rangle =0$ is given by%
\begin{equation}
f(x)=\frac{c+dx}{a+bx},  \label{cv}
\end{equation}%
where $a,b,c,d$ are real constants. In a space--time region where the
function $f(x)$ takes the form of Eq.\ (\ref{cv}), \ there is no difference
between the state $|\Phi \rangle $ and the vacuum $|0\rangle $ about quantum
field fluctuation, which will be coupled to and excited by the switch body
in the later analysis. Due to the residual gauge symmetry, the fundamental
many-point functions of the input states are not $\langle \Phi |\phi
_{+}(x_{1}^{+})\cdots \phi _{+}(x_{N}^{+})|\Phi \rangle $, but are instead $%
\langle \Phi |\dot{\phi}_{+}(x_{1}^{+})\cdots \dot{\phi}_{+}(x_{N}^{+})|\Phi
\rangle $. First of all, it can be easily checked that both the one-point
functions of $\dot{\phi}_{+}$ vanish: $\langle \Phi |\dot{\phi}%
_{+}(x_{1}^{+})|\Phi \rangle =\langle 0|\dot{\phi}_{+}(x_{1}^{+})|0\rangle
=0 $. The two-point function of $|\Phi \rangle $ is calculated as%
\begin{equation}
\langle \Phi |\dot{\phi}_{+}(x_{1}^{+})\dot{\phi}_{+}(x_{2}^{+})|\Phi
\rangle =-\frac{\hbar }{4\pi }\frac{\dot{f}(x_{1}^{+})\dot{f}(x_{2}^{+})}{%
\left( f(x_{1}^{+})-f(x_{2}^{+})\right) ^{2}}.  \label{2p}
\end{equation}%
Substituting eq(\ref{cv}) into eq(\ref{2p}), it is possible to show by
explicit calculations that 
\begin{equation}
\langle \Phi |\dot{\phi}_{+}(x_{1}^{+})\dot{\phi}_{+}(x_{2}^{+})|\Phi
\rangle =\langle 0|\dot{\phi}_{+}(x_{1}^{+})\dot{\phi}_{+}(x_{2}^{+})|0%
\rangle .  \label{conformal}
\end{equation}%
Note that the many-point functions of in-field, which describe the initial
conditions of the system, can be decomposed via the Wick's theorem into a
sum of the products of the two-point functions of the state. Therefore, when
the input state is $|\Phi \rangle $, all the local properties in a
space--time region where Eq.\ (\ref{cv}) \ holds are equal to those of the
vacuum input. As an instructive example, let us consider a situation in
Figure 1. For the state $|\Phi \rangle $, local quantum fluctuation of a
spacetime region $R$ in which the energy flux vanishes ($\langle \Phi
|T_{\mu \nu }|\Phi \rangle =0$) are the same as that of the vacuum state $%
|0\rangle $. Hence any local event $P$ which takes place in $R$ evolves as
if the initial state were the vacuum state $|0\rangle $ (Figure 2), as far
as the sequential events originated from event $P$ happen in the region $R$.
This property comes from the conformal \ symmetry of the massless field and
is very crucial to derive the lower bound of actuating energy of photon
switching in the later discussion.

\bigskip

\section{ Universal lower bound of actuating energy derived by a gedanken
experiment}

\ \newline

We consider now the switching process. A perfect mirror boundary of the
switch is located at $x=0$ in the $yz$ plane. The mirror body lies on the
left-hand side of the boundary. Until $t=0$, the mirror reflects incident
left-moving beams with arbitrary shapes arriving at $x=0$ . From $t=0$, the
mirror and its body gradually become transparent. During the switching
interval, the switch generates undesired photons, which energy gives a lower
bound of $E_{s}$. Incident beams arriving at $x=0$ after $t=t_{s}$ penetrate
completely and continue to propagate freely in the spatial region $x\leq 0$.

\bigskip

By adopting not Schr\H{o}dinger's but Heisenberg's picture of operators and
states, the quantum field can be described by both in- and out- asymptotic
fields. In the infinite past, the in-asymptotic field $\phi _{in}~$evolves
freely and in the infinite future the out-asymptotic field $\phi _{out}$
evolves freely. In the past before the mirror begins to be transparent,
plane-wave mode functions of $\phi _{in}$ are given by%
\begin{equation*}
U_{\omega }^{in}\left( t,x\right) =\Theta (x)\left[ u_{\omega
}(x^{+})-u_{\omega }(x^{-})\right] ,
\end{equation*}%
where $u_{\omega }$ is defined by Eq.(\ref{p-mode}). Note that the mirror
boundary condition is satisfied as follows.%
\begin{equation*}
U_{\omega }^{in}(t,0)=0.
\end{equation*}%
In the past region, $\phi _{in}$ is expanded as

\begin{equation*}
\phi _{in}(t,x)=\int_{0}^{\infty }d\omega \left[ A_{\omega }^{in}U_{\omega
}+A_{\omega }^{in\dag }U_{\omega }^{\ast }\right] ,
\end{equation*}%
where $A_{\omega }^{in},~$ $A_{\omega }^{in\dag }$ are annihilation and
creation operators of the in-particles. The in -vacuum state is defined by $%
A_{\omega }^{in}|0,in\rangle =0.$ The initial state is $|0,in\rangle $ for
the switching process. In order to solve the scattering problem by the
mirror for two shock waves with negative and positive energy density, let us
introduce another set of mode functions, which satisfies the mirror boundary
condition. For an arbitrary monotonically increasing $C^{1}~$function $f(x)$%
of $x\in \left( -\infty ,\infty \right) $satisfying $f(\pm \infty )=\pm
\infty $, the mode functions are given by

\begin{equation}
V_{\omega }^{in}\left( t,x\right) =\Theta (x)\left[ v_{\omega
}(x^{+})-v_{\omega }(x^{-})\right] ,  \label{vinm}
\end{equation}%
where $v_{\omega }$ is given by Eq.(\ref{vm}) and $V_{\omega }^{in}\left(
t,0\right) =0$. It is possible to expand $\phi _{in}$ as 
\begin{equation*}
\phi _{in}(t,x)=\int_{0}^{\infty }d\omega \left[ B_{\omega }^{in}V_{\omega
}+B_{\omega }^{in\dag }V_{\omega }^{\ast }\right] ,
\end{equation*}%
A squeezed in-state $|\Phi ,in\rangle $ is defined by $B_{\omega }^{in}|\Phi
,in\rangle =0.~$If we take the same function $f$ as Eq.(\ref{f3}), $|\Phi
,in\rangle $ gives the expectation value of the in-coming energy flux \ in
Eq.(\ref{n+p}). If $x_{i}<0$, the negative flux in Eq.(\ref{n+p}) is firstly
reflected to the right direction by the mirror and propagating freely. As
seen in Eq.(\ref{vinm}), the mirror does \ not make any tail of the
wavepacket and the negative energy flux keeps its shock-wave shape. The
switching process of the mirror after $t=0$ does not affect the evolution of
the wavepacket with negative energy because the wave runs away at light
velocity and causality of the system prevent the switching event from
disturbing the wavepacket's evolution. Hence even in the remote future the
negative flux propagates to the right with its localized shape and is
spatially separated from positive-energy right-moving excitations. Here a
comment is added that if spatial support of the mirror is finite and given
by $\left[ x_{m},~0\right] $, then we need another set of in-mode functions
given by

\begin{equation*}
U_{\omega }^{\prime in}\left( t,x\right) =\Theta (x_{m}-x)\left[ u_{\omega
}(x^{-}-x_{m})-u_{\omega }(x^{+}+x_{m})\right] ,
\end{equation*}%
which satisfy the boundary conditions $U_{\omega }^{\prime in}\left(
t,x_{m}\right) =0$. Then $\phi _{in}$ is expanded as 
\begin{eqnarray*}
\phi _{in}(t,x) &=&\int_{0}^{\infty }d\omega \left[ A_{\omega
}^{in}U_{\omega }+A_{\omega }^{in\dag }U_{\omega }^{\ast }\right] \\
&&+\int_{0}^{\infty }d\omega \left[ A_{\omega }^{\prime in}U_{\omega
}^{\prime }+A_{\omega }^{\prime in\dag }U_{\omega }^{\prime \ast }\right] .
\end{eqnarray*}%
The in-vacuum state is redefined by $A_{\omega }^{in}|0,in\rangle =A_{\omega
}^{\prime in}|0,in\rangle =0$, and the squeezed in-state $|\Phi ,in\rangle $
is redefined by $B_{\omega }^{in}|\Phi ,in\rangle =A_{\omega }^{\prime
in}|\Phi ,in\rangle =0$.

\bigskip

In the remote future region after the mirror has been completely removed,
two sets of free plane-wave mode functions expands the out-asymptotic field $%
\phi _{out}$ as follows.

\begin{eqnarray*}
\phi _{out}(t,x) &=&\phi _{+}^{out}\left( x^{+}\right) +\phi
_{-}^{out}\left( x^{-}\right) \\
&=&\int_{0}^{\infty }d\omega \left[ A_{\omega }^{+out}u_{\omega
}(x^{+})+A_{\omega }^{+out\dag }u_{\omega }^{\ast }(x^{+})\right] \\
&&+\int_{0}^{\infty }d\omega \left[ A_{\omega }^{-out}u_{\omega
}(x^{-})+A_{\omega }^{-out\dag }u_{\omega }^{\ast }(x^{-})\right] ,
\end{eqnarray*}%
where $A_{\omega }^{\pm out\dag }$ and $A_{\omega }^{\pm out}$ are creation
and annihilation operators of the out-particles. The out-vacuum state is
defined by $A_{\omega }^{\pm out}|0,out\rangle =0.$ It should be stressed
that the energy-momentum tensor operators is normal-ordered so as to its
expectational value for $|0,out\rangle $ vanishes. Hence, the state $%
|0\rangle $ in Eq.(\ref{ps}) is regarded as $|0,out\rangle $.

\bigskip

Let us consider a gedanken experiment in order to evaluate the inevitable
work for the switch to produce undesired photons. Instead of the vacuum
state $|in,0\rangle $, we take as an input quantum state the squeezed state $%
|\Phi ,in\rangle $ \ which satisfies Eq.\ (\ref{n+p}) with $x_{i}\leq 0$ and 
$x_{f}\geq t_{s}$ . The negative flux in Eq.\ (\ref{n+p}) reaches $x=0$ when 
$t=x_{i}$ and is perfectly reflected. From $t=x_{i}+0$ to $t=0\,$, no input
or output beams reach the switch body. From $t=0$, the mirror boundary and
its body gradually become transparent. The switch generates undesired
photons, which energy $E_{\Phi }$ gives a lower bound of actuating energy $%
E_{s,\Phi }$ of the switch in the case. When the positive flux in Eq.\ (\ref%
{n+p}) reaches $x=0$ at $t=x_{f}(\geq t_{s})$, the mirror has been already
removed at the point and the flux propagates freely to the left. After $%
t=t_{s}$, the right-moving flux of the field is described as 
\begin{equation}
\langle \Phi ,in|T_{--}(x^{-})|\Phi ,in\rangle =-\left\vert E_{n}\right\vert
\delta (x^{-}-x_{i})+\Theta \left( x^{-}\right) \Theta \left(
t_{s}-x^{-}\right) T_{s}(x^{-}),
\end{equation}%
where $T_{s}(x^{-})$ denotes the undesired flux induced by the switching.
For the total energy of the undesired flux $E_{\Phi
}=\int_{0}^{t_{s}}T_{s}(x^{-})dx^{-}$, \ the following inequality holds:

\begin{equation}
E_{\Phi }\geq \frac{\left\vert E_{n}\right\vert }{1-\frac{12\pi }{\hbar }%
\left\vert E_{n}\right\vert |x_{i}|},  \label{ieq1}
\end{equation}%
as will be proven later. Because no input or output energy flux reaches the
mirror body from $t=x_{i}+0$ to $t=0$ and Eq.\ (\ref{cv}) is satisfied
around the mirror body during the interval $|x_{i}|$, the conformal
asymmetry argument mentioned above can apply. We are able to regard
production and evolution of the undesired photons as event P and its
sequential events of $R$ in Figure 1. Consequently, the actuating energy $%
E_{s,\Phi }$ of the switch is independent of $|\Phi ,in\rangle $ \ and
exactly equal to the actuating energy $E_{s}$ of the switch with the input
vacuum state $|0,in\rangle $. Therefore $E_{s}$, which we are interested in,
is lower-bounded by the undesired photon energy $E_{\Phi }$ of the state $%
|\Phi ,in\rangle $. \ It should be noted that the inequality of Eq.\ (\ref%
{ieq1}) is satisfied by all possible input fluxes in Eq.\ (\ref{n+p}).
Therefore, the inequality

\begin{equation}
E_{s}\geq \max_{\left\{ |E_{n}|,~x_{f},~x_{i}\right\} }\frac{\left\vert
E_{n}\right\vert }{1-\frac{12\pi }{\hbar }\left\vert E_{n}\right\vert |x_{i}|%
}  \label{ieq2}
\end{equation}%
must hold. Here $x_{f}$, $x_{i}$ and $|E_{n}|$ are constants which appear in
Eq.\ (\ref{n+p}). For fixed $x_{f}$ and $~|E_{n}|$, maximizing $\frac{%
\left\vert E_{n}\right\vert }{1-\frac{12\pi }{\hbar }\left\vert
E_{n}\right\vert |x_{i}|}$ is achieved by maximizing $|x_{i}|$. From the
uncertainty relation in Eq.\ (\ref{nieq}), $|x_{i}|$ is bounded above as 
\begin{equation}
|x_{i}|=\left\vert x_{f}-x_{i}\right\vert -|x_{f}|\leq \frac{\hbar }{12\pi
|E_{n}|}-|x_{f}|.
\end{equation}%
Thus the maximum of $|x_{i}|$ is $\frac{\hbar }{12\pi |E_{n}|}-|x_{f}|$.
Substituting the maximum value into $\frac{\left\vert E_{n}\right\vert }{1-%
\frac{12\pi }{\hbar }\left\vert E_{n}\right\vert |x_{i}|}$, we obtain 
\begin{equation}
E_{s}\geq \max_{\left\{ x_{f},~|E_{n}|\right\} }\frac{\hbar }{12\pi |x_{f}|}.
\end{equation}%
Because\bigskip $\frac{\hbar }{12\pi |x_{f}|}$ does not depend on $|E_{n}|$,
the inequality becomes 
\begin{equation}
E_{s}\geq \max_{x_{f}}\frac{\hbar }{12\pi |x_{f}|}.
\end{equation}%
Note that the minimum of $|x_{f}|$ is $t_{s}$. Hence we obtain a result,
ignoring photon polarization:%
\begin{equation}
E_{s}\geq \frac{\hbar }{12\pi t_{s}}.  \label{wieq}
\end{equation}%
In the real situations, photon beams have two massless fields corresponding
to two polarizations. Hence, the lower bound of $E_{s}$ should be doubled:%
\begin{equation}
E_{s}\geq \frac{\hbar }{6\pi t_{s}}.  \label{wieq1}
\end{equation}%
This result is our main result. The bound holds for the inverse operation to
switch perfect transmission to perfect reflection. Because the photon switch
can be coupled to other electromagnetic modes propagating to directions
different from\ the $x$ axis, the switch may generate undesired photons in
the other modes. We have also neglected the contribution of undesired
left-moving excitations which might be generated during the switching.
However, even if these additional contributions exist, the bound of Eq.\ (%
\ref{wieq1}) remains valid because the contributions just increase $E_{s}$.
Therefore we can conclude that the bound in Eq.(\ref{wieq1}) is universal.

\bigskip

\bigskip

\section{Inevitable work to produce undesired photons}

\bigskip

We now consider the inequality in Eq.\ (\ref{ieq1}). It is enough to discuss
the out- asymptotic region in the remote future, where the negative flux and
additional flux produced by the mirror switch make free evolution. Thus no
need to take account of any boundary conditions in the whole space $\left(
-\infty ,\infty \right) $. This inequality comes from a fact that if we are
given a negative-energy flux in some spatial region, there must exist
positive-energy flux in other regions in order to guarantee the positivity
of total energy in the whole space. Let us take an arbitrary out-state $%
|\Psi ,out\rangle $ of $\phi _{-}\left( x^{-}\right) $ which satisfies%
\begin{equation}
\langle \Psi ,out|T_{--}(x^{-})|\Psi ,out\rangle =-\left\vert
E_{n}\right\vert \delta (x^{-})  \label{t1}
\end{equation}%
for $-\infty <x^{-}<L$. Here, the positive constant $\left\vert
E_{n}\right\vert $ is fixed independent of $|\Psi ,out\rangle $. The states $%
|\Psi ,out\rangle $ are not restricted to in-squeezed states. \ Eq.\ (\ref%
{t1}) can be expressed using $x^{-}\in \left( -\infty ,\infty \right) $ as 
\begin{equation}
\langle \Psi ,out|T_{--}(x^{-})|\Psi ,out\rangle =-\left\vert
E_{n}\right\vert \delta (x^{-})+\Theta \left( x^{-}-L\right) T_{s}(x^{-}-L),
\end{equation}%
where $T_{s}(x^{-}-L)$ depends on the details of the state $|\Psi
,out\rangle $ and is constrained such that $E=\int_{-\infty }^{\infty
}\langle \Psi ,out|T_{--}(x)|\Psi ,out\rangle dx$ is positive. We use the
Lagrange multiplier method to find a quantum state $|\Psi ,out\rangle $
which minimizes the total energy $E$ with a constraint in Eq.\ (\ref{t1}).
This requires minimizing the quantity $I$, defined by\bigskip

\begin{align*}
I& =\int_{-\infty }^{\infty }\langle \Psi ,out|T_{--}(x)|\Psi ,out\rangle
dx+\int_{-\infty }^{L}\eta (x)\left[ \langle \Psi ,out|T_{--}(x)|\Psi
,out\rangle +\left\vert E_{n}\right\vert \delta (x)\right] dx \\
& +\lambda \left[ \langle \Psi ,out|\Psi ,out\rangle -1\right] ,
\end{align*}%
where $\eta (x)$ and $\lambda $ are multipliers. Because we do not need any
conditions on the energy flux at $x=L$, we set $\eta (L)=0$. From the
variation of $I$ in terms of $\lambda $, we get the normalization condition
of the state $|\Psi ,out\rangle $. From the variation of $I$ in terms of $%
\eta (x)$, Eq.\ (\ref{t1}) is reproduced. The variation of $I$ in terms of $%
|\Psi ,out\rangle $ leads to the eigenvalue equation 
\begin{equation}
F|\Psi ,out\rangle =-\lambda |\Psi ,out\rangle ,
\end{equation}%
where $F$ is a Hermitian operator defined by

\begin{equation}
F=\int_{-\infty }^{\infty }\left[ 1+\eta (x)\Theta \left( L-x\right) \right]
T_{--}(x)dx.
\end{equation}%
The operator $F$ can be diagonalized by a method given by Flanagan \cite{F}.
We define a $C^{1}$ function $f(x)$ such that 
\begin{equation}
f(x)=x
\end{equation}
for $x\in (L,\infty )~$and 
\begin{equation}
f(x)=L-\int_{x}^{L}\frac{du}{1+\eta (u)}
\end{equation}
for $x\in (-\infty ,L)$. Assuming a monotonic increase of $f(x)$ and $%
f(-\infty )=-\infty $, which will be justified later, we can expand the
right-moving field $\phi _{-}$ such that%
\begin{equation*}
\phi _{-}\left( x^{-}\right) =\int_{0}^{\infty }d\omega \sqrt{\frac{\hbar }{%
4\pi \omega }}\left[ b_{\omega }^{R}e^{-i\omega f(x^{-})}+b_{\omega }^{R\dag
}e^{i\omega f(x^{-})}\right] ,
\end{equation*}%
where $b_{\omega }^{R\dag },~b_{\omega }^{R}$ are creation and annihilation
operators. The operator $F$ can be rewritten \cite{F} as

\begin{equation}
F=\int_{0}^{\infty }\hbar \omega b_{\omega }^{R\dag }b_{\omega }^{R}d\omega -%
\frac{\hbar }{12\pi }\int_{-\infty }^{\infty }\left( \partial _{x}\sqrt{%
1+\eta (x)\Theta \left( L-x\right) }\right) ^{2}dx.
\end{equation}%
For a fixed $\eta (x)$, the normalized eigenstate $|vac,\eta \rangle $ with
the minimum eigenvalue of $F$ satisfies $b_{\omega }^{R}|vac,\eta \rangle
=0. $ This simplifies the problem because the state $|vac,\eta \rangle $ is
a squeezed state in the meaning mentioned above. Its generating function is
denoted by $f_{\eta }(x)$. We can find a function $f_{\eta }(x)$ which
generates plain-wave mode functions for both $x>L(>0)$ and $x<0$ and
satisfies Eq.\ (\ref{t1}), as follows:

\begin{align*}
f_{\eta }(x)& =\Theta (x-L)x \\
& +\Theta (L-x)\Theta (x)\left[ \frac{1}{\rho ^{2}\left( L-x\right) +\rho }-%
\frac{1}{\rho }+L\right] \\
& +\Theta (-x)\left[ \frac{x}{\left( \rho L+1\right) ^{2}}+\frac{1}{\rho
^{2}L+\rho }-\frac{1}{\rho }+L\right] ,
\end{align*}%
where the constant $\rho $ is defined by $\rho =\frac{\left\vert
E_{n}\right\vert }{\frac{\hbar }{12\pi }-\left\vert E_{n}\right\vert L}$. It
is easy to check that this function $f_{\eta }(x)$ is a monotonically
increasing $C^{1}$ function with $f_{\eta }(\pm \infty )=\pm \infty $. The
corresponding squeezed state $|vac,\eta \rangle $ takes a minimum of $E$.
The energy flux of $|vac,\eta \rangle $ is calculated as

\begin{equation}
\langle vac,\eta |T_{--}(x^{-})|vac,\eta \rangle =-\left\vert
E_{n}\right\vert \delta (x^{-})+\frac{\left\vert E_{n}\right\vert }{1-\frac{%
12\pi }{\hbar }\left\vert E_{n}\right\vert L}\delta \left( x^{-}-L\right) .
\label{ev}
\end{equation}%
By integrating the second term in the right-hand-side of Eq.\ (\ref{ev}), it
is shown that the minimum of $E_{\Phi }$ is given by $\frac{\left\vert
E_{n}\right\vert }{1-\frac{12\pi }{\hbar }\left\vert E_{n}\right\vert L}$.
By substituting $L=|x_{i}|$, Eq.\ (\ref{ieq1}) is exactly derived.

\bigskip

\textbf{Acknowledgments}\newline

\bigskip

I would like to thank M. Ozawa, M. Morikawa and A. Shimizu for useful
discussions. This research was partially supported by the SCOPE project of
the MIC.

\bigskip

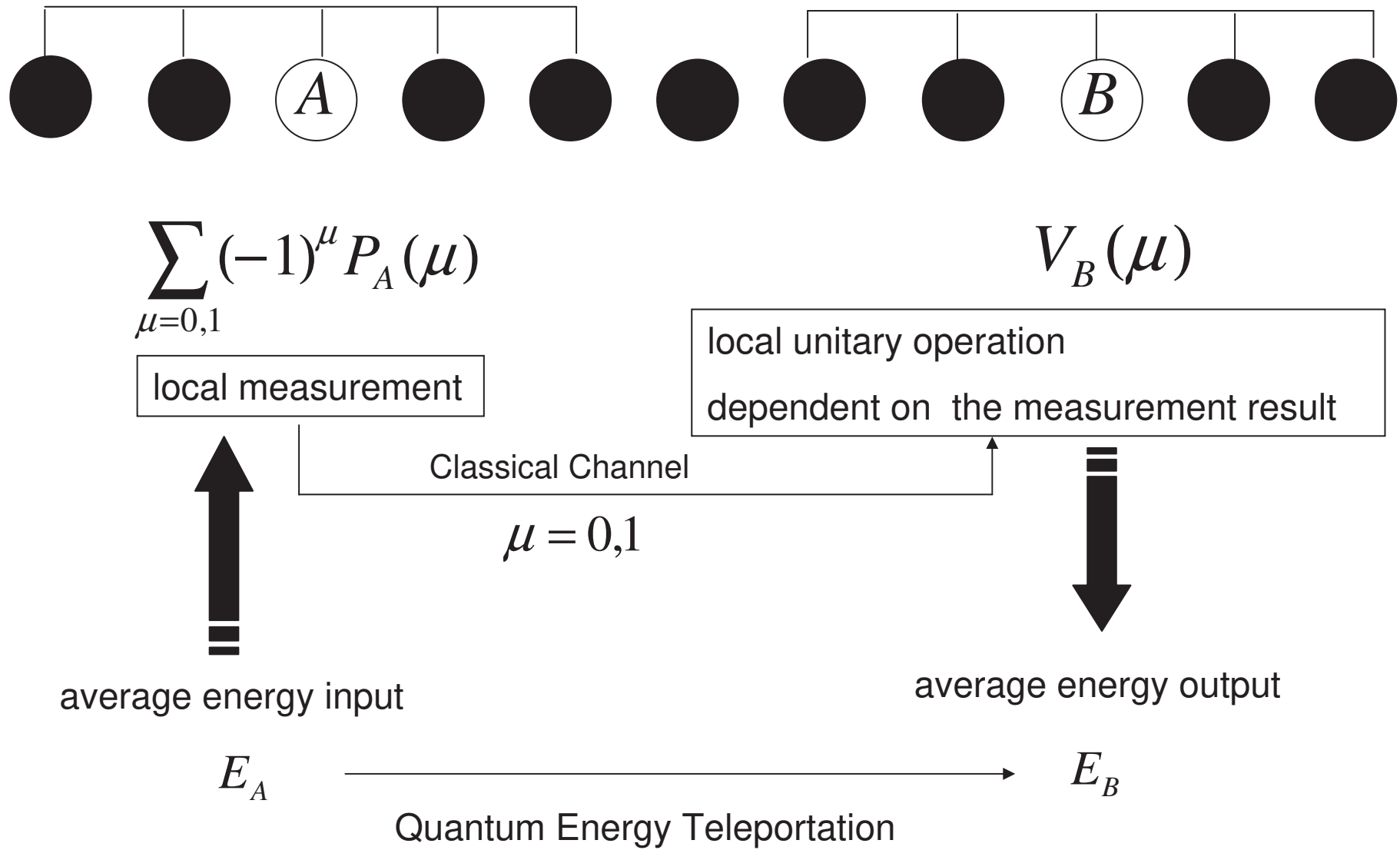
Figure 1: For the state $|\Phi \rangle $, local quantum fluctuation of a
spacetime region $R$ in which the energy flux vanishes ($\langle \Phi
|T_{\mu \nu }|\Phi \rangle =0$) are the same as that of the vacuum state $%
|0\rangle $. Hence any local event $P$ which takes place in $R$ evolves as
if the initial state were the vacuum state $|0\rangle $, as far as the
sequential events originated from $P$ happen in the region $R$.

\bigskip

Figure 2: $P$ in the state $|0\rangle $ evolves in the same way of the $%
|\Phi \rangle $ case, as far as the sequential events originated from $P$
happen in the same region of $R$ of Figure 1.

\bigskip

\end{document}